**Mining of high throughput screening database reveals AP-1 and autophagy pathways as potential targets for COVID-19 therapeutics**

Hu Zhu,[1] Catherine Z. Chen[1], Srilatha Sakamuru[1], Anton Simeonov,[1] Mathew D. Hall,[1] Menghang Xia[1], Wei Zheng[1], Ruili Huang[1*]

[1]Division of Preclinical Innovation, National Center for Advancing Translational Sciences (NCATS), National Institutes of Health (NIH), Rockville, MD 20850, USA.

*Address correspondence and reprint requests to

Ruili Huang, Ph.D.

9800 Medical Center Drive

DPI/NCATS

National Institutes of Health

Rockville, MD 20850

Phone: 301-827-0944

Fax: 301-217-5736

Email: huangru@mail.nih.gov

**Abstract**

The recent global pandemic of Coronavirus Disease 2019 (COVID-19) caused by the new coronavirus SARS-CoV-2 presents an urgent need for new therapeutic candidates. Many efforts have been devoted to screening existing drug libraries with the hope to repurpose approved drugs as potential treatments for COVID-19. However, the antiviral mechanisms of action for the drugs found active in these phenotypic screens are largely unknown. To deconvolute the viral targets for more effective anti-COVID-19 drug development, we mined our in-house database of approved drug screens against 994 assays and compared their activity profiles with the drug activity profile in a cytopathic effect (CPE) assay of SARS-CoV-2. We found that the autophagy and AP-1 signaling pathway activity profiles are significantly correlated with the anti-SARS-CoV-2 activity profile. In addition, a class of neurology/psychiatry drugs was found significantly enriched with anti-SARS-CoV-2 activity. Taken together, these results have provided new insights into SARS-CoV-2 infection and potential targets for COVID-19 therapeutics.

## Introduction

In late fall 2019, a new Coronavirus Disease 2019 (COVID-19) emerged from Wuhan, China. It is caused by severe acute respiratory syndrome coronavirus 2 (SARS-CoV-2). SARS-CoV-2 appears to be highly contagious, and a lack of immunity in the human population has resulted in rapid spread across the globe. As of June 28rd, 2020, it has infected over 10 million people, killed over 500,000 people, and caused abrupt disruption of social and economic activity across the world (https://covid19.who.int/).

Currently, there is no effective treatment for COVID-19. Drug development typically takes 12-16 years and costs US$1-2 billion to bring a new drug to market[1]. Preventative approaches such as vaccines and antibodies could also take years to develop. Given that treatments for patients infected with SARS-CoV-2 are needed immediately, repurposing existing drugs and clinical investigational drugs to treat COVID-19 is an attractive strategy. This approach takes advantage of known human pharmacokinetic and safety profiles of drugs, which allow rapid initiation of human clinical trials or direct use for treatments. Remdesivir is a good example of such effort to treat COVID-19. Remdesivir was originally developed for RNA viruses and was then tested in a clinical trial against Ebola virus infection during the 2016 outbreak [2,3]. After remdesivir was shown to be active against SARS-CoV-2 *in vitro*[4], clinical trials have been rapidly conducted. In a double-blind, randomized, placebo-controlled trial carried out by the National Institutes of Health (NIH), remdesivir was demonstrated effective in reducing the recovery time from 15 days to 11 days in hospitalized COVID-19 patients[5]. On May 1st, 2020, the U.S. Food and Drug Administration (FDA) issued an emergency use authorization (EUA) for the investigational antiviral drug remdesivir for the treatment of hospitalized COVID-19 patients. While many clinical trials are still ongoing and are showing promising results (https://clinicaltrials.gov/ct2/results?cond=COVID-19), some repurposing efforts have had disappointing or controversial outcomes, for example, those for lopinavir-ritonavir[6], hydroxychloroquine (HCQ), and chloroquine (CQ)[7,8].

Other than an intuitive repurposing approach based on a known mechanism (such as the recent positive reports of dexamethasone for modulating inflammatory response in COVID19)[9], an unbiased and systematic screening of approved drug or clinical investigational drugs might discover additional therapeutic options. Multiple sites [10-14], including our center (The National Center for Advancing Translational Sciences, NCATS), are screening approved drug and mechanistically annotated libraries to identify new therapeutics. To rapidly share the screening results and accelerate the drug repurposing process, NCATS has posted all screening data on an online database (Open Science Data Portal of COVID-

19) (https://opendata.ncats.nih.gov/covid19/index.html) that is freely available to the public[15]. In most antiviral drug repurposing efforts, the most scalable assay used for screening in biological safety level-3 laboratories is a phenotypic assay, measuring the cytopathic effect (CPE) of SARS-CoV-2 virus on Vero E6 cells infected for 72 hours. If compounds have antiviral activity, Vero E6 cells are rescued from the CPE. While there are many drugs with known targets/mechanisms of action for their approved indications, the targets or mechanisms of their antiviral activity are largely unknown, be it against a host of viral target [10-14]. It is thus crucial to better understand their antiviral mechanisms to facilitate further drug development.

The NCATS Pharmaceutical Collection (NPC) [16] is a library of ~3,000 drugs approved for marketing in the US (FDA), Europe (EMA), Canada, Australia, and/or Japan (PMDA). The library was specifically created to enable drug repurposing and has been screened at NCATS in nearly 1,000 assays in concentration-response (quantitative high throughput screening, qHTS), encompassing a wide range of disease targets and pathways with main disease areas covered including rare and neglected diseases, infectious diseases and cancer. Here, we leveraged this unique dataset to compare activity across SARS-CoV-2 CPE screening data (both from NCATS and published elsewhere) [6,14,17-21] with historical in house NPC qHTS data. Correlations were performed to identify assays with patterns of activity similar to that of the SARS-CoV-2 CPE assay.

## Results

### *Mining qHTS data reveals interesting anti-SARS-CoV-2 targets: Autophagy and AP-1 signaling*

Screening compounds using phenotypic assays, such as the CPE assay, has identified compounds that inhibited cell death caused by the SARS-CoV-2 infection. Comparing the NPC compound activity profile in each of the ~1000 screens performed previously on various targets with the activity profile against SARS-CoV-2 may help identify targets of the compounds with anti-SARS-CoV-2 activity, and provide important clues to the underlying targets and mechanisms of the pathogenesis of SARS-CoV-2. Assays with activity profiles which resemble that of SARS-CoV-2 could serve as targets for the development of new COVID-19 therapies. Toward this goal, we collected compounds reported as active from recent anti-COVID-19 repurposing screens using the SARS-CoV-2 CPE assay[14,17,18] and drugs proposed by the scientific community as potential COVID-19 therapies.[10,19-21] Activities of these compounds were used as a “probe signature” to compare with the activity profiles of all other assays (**Figure 1(a)**). Compound activity was represented by "curve rank",[22,23] a numeric measure between -9 and 9 based on potency, efficacy, and the quality of the concentration response curve, such that a large positive number indicates a strong activator, a large negative number indicates a strong inhibitor, and 0 means inactive. Activity profile similarity was measured by the Pearson Correlation Coefficient (*r*) with a p-value calculated for the significance of correlation (**Figure 2**).

The activity profiles from an autophagy assay ($r = 0.47$, $p<1\times10^{-20}$)[24] and an AP-1 signaling pathway assay ($r = 0.37$, $p<1\times10^{-20}$)[25] exhibited the most significant correlations with that of the SARS-CoV-2 screen **(Figure 1(b))**. Interestingly, two other antiviral assays, an Ebola virus-like particle entry assay (EBOV) ($r = 0.39$, $p<1\times10^{-20}$)[26] and a MERS pseudo particle entry assay ($r = 0.28$, $p<1\times10^{-20}$),[27] were also among the most significantly correlated assays with activity profiles that highly resemble that of the SARS-CoV-2 CPE assay **(Figure 1(b))**. As MERS belongs to the same family of beta-coronaviruses as SARS-CoV-2, this finding can serve as a validation of our approach.[28-30] Moreover, remdesivir, an antiviral drug active against Ebola was recently found effective against SARS-CoV-2 in *vitro* assays and approved for treating hospitalized COVID-19 patients.[5,31] This is consistent with our finding that a significant number (118) of drugs including remdesivir that showed anti-Ebola activity also showed activity in the SARS-CoV-2 CPE assay, suggesting some shared drug targets (either viral target or cellular targets) between EBOV and SARS-CoV-2autophagy and AP-1 assays were combined, i.e., a compound was counted as active if it was active in either one of these assays and inactive otherwise, the sensitivity in picking up SARS-CoV-2 actives increased to 0.85 with an improved BA of 0.81. This suggests that the targets in these two assays are different and either autophagy or AP-1 could only account for one mechanism in targeting SARS-CoV-2, thus combining the two pathways might increase the likelihood of identifying drugs that could target SARS-CoV-2 through different mechanisms.

A list of the most potent compounds (<20 µM) in the AP-1 assay and their corresponding activities in the autophagy and SARS-CoV-2 CPE assays are provided in **Table 2**. These drugs could be considered for further anti-COVID-19 development. Concentration-response curves of exemplar compounds that were active in all three assays are shown in **Figure 3**.

*Enrichment of neuroactive drugs in anti-SARS-CoV-2 and AP-1 active compounds*

Another interesting phenomenon we observed is that a large number of compounds active in the SARS-CoV-2 CPE assay are psychoactive drugs.  We next investigated the statistical significance of this finding. There were 359 drugs annotated as neurology/psychiatry drugs tested in the SARS-CoV-2 CPE assay. We found that 74 of them are active (21%) (**Supplementary Table 1**), whereas only 8% of the drugs not in this category were active in the SARS-CoV-2 CPE assay, corresponding to a 2.6-fold enrichment of actives in the neuroactive drugs. This enrichment is statistically significant (Fisher's exact test: $p = 2.41\times10^{-11}$). To check whether this phenomenon only occurs in this class of drugs, we also examined five other common drugs classes, including infectious disease, cardiology, endocrinology, gastroenterology, and oncology. We found that none of these classes were significantly enriched with the anti-SARS-CoV-2 active compounds (**Figure 4**). The results suggest possible connections between the psychoactive drugs and the targets/pathways related to SARS-CoV-2 infection or replication in host cells.

## Discussion

To deconvolute the viral targets for more effective anti-COVID-19 drug development, we compared the compound activity profiles from our historical qHTS data with recent SARS-CoV-2 CPE assay data. We found that activities against autophagy and AP-1 significantly correlated with anti-SARS-CoV-2 activity. We also found strong correlations between SARS-CoV-2 and other antiviral assays such as MERS-CoV pseudo particle entry. It is intuitive given that both are zoonotic beta-coronaviruses with similar genomes and a common cellular entry mechanism. Since the identification of SARS-CoV-2, multiple agents shown to be active against MERS-CoV and SARS-CoV over the past 15 years have been tested and demonstrated to retain activity against SARS-CoV-2. The analysis here of independent assays performed years apart reinforces this observation.

Autophagy and endocytosis are interconnected cellular pathways for the degradation and recycling of intracellular and extracellular components, respectively. The two pathways interact and interdepend on each other, sharing some molecular machinery[32]. The autophagy/endocytosis pathway has been implicated in the entry of coronavirus into host cells, including SARS-CoV, MERS-CoV and SARS-CoV-2[12,33]. In our recent study, a small number of autophagy modulators were tested to clarify whether the activity of CQ/HCQ was related to its autophagy modulatory properties, and a number of active compounds were confirmed in the CPE assay[12]. Here, our unbiased comparison of the SARS-CoV-2 CPE assay with approximately 1,000 NCATS qHTS assays targeting various drugs targets and diseases found a significant correlation with an autophagy assay screened against the NPC library several years earlier, further validating our data mining approach. Targeting the autophagy pathway has been tested in clinical trials for curbing COVID-19. For example, CQ and HCQ are antimalaria drugs and known autophagy inhibitors. HCQ/CQ have shown promising anti-SARS-CoV-2 activity *in vitro*[12,34,35]; however, their therapeutic effect in COVID-19 patients are still controversial[7,8]. Our analysis here reinforces that more selective and potent modulators of autophagy pathways should be further evaluated in pre-clinical models for antiviral activity. Another factor that needs to be taken into consideration is that the coronavirus can take two distinct pathways for cell entry, either endosomal or non-endosomal. Blocking autophagy, which is the endocytosis pathway, might not be sufficient to block the viral entry, and combination approaches, e.g., combination treatments with autophagy inhibitors and TMPRSS2 inhibitors[36], warrant consideration.

Activator protein 1 (AP-1) is a dimeric transcription factor composed of proteins belonging to the Jun, Fos, ATF and JDP families and regulates a range of cellular processes. The AP-1 transcription factor family could be activated by different stimuli, such as cytokines, stress, bacterial and viral infections[37]. The AP-1 signaling pathway has been shown to be activated by the SARS-CoV viral particle[38], the spike protein[39], the nucleocapsid protein[40] and the accessory protein 3b[41]. In a recent study, Jun, one of the AP-1 proteins, has been identified as one of the top hub host proteins, which is directly targeted by CoV proteins or indirectly involved in the CoV infection [42]. The activation of AP-1 signaling might serve as an immune response for the host to fight viral infections. One hypothesis that can be drawn from this observation is that the AP-1 pathway may be hijacked by the coronavirus and mediate the process of the CPE, and disruption of the AP-1 pathway could offset this process. While this hypothesis has not been directly tested, the correlation we found between AP-1 and SARS-CoV-2 points to this as a druggable host pathway for SARS-CoV-2 and future emergent coronaviruses.

Psychoactive drugs have been reported to be active against SARS-CoV-2[10,14,17,18]. Here we found that the neurology/psychiatry class of drugs, in contrast to other classes of drugs, was significantly enriched in anti-SARS-CoV-2 activity. Most of the active compounds in the neurology/psychiatry class of drugs are psychoactive drugs, which target G protein coupled receptors (GPCRs), particularly monoamine receptors (86% of the psychoactive drugs that showed anti-SARS-CoV-2 activity belong to this category) (**Supplementary Table 1**). We hypothesize that those compounds might bind to membrane receptors and activate intracellular pathways to fight coronaviruses. It is interesting that among the compounds that were active in both the AP-1 and SARS-CoV-2 CPE assays, we found a more pronounced enrichment of neurology/psychiatry drugs (3.76-fold; $p = 3.89\times10^{-11}$), suggesting that these drugs may also act through the AP-1 pathway to inhibit SARS-CoV-2. Another hypothesis is that SARS-CoV-2 might infect cells through other unknown membrane proteins in addition to angiotensin-converting enzyme 2 (ACE2) and those compounds might interfere with the viral binding to its receptors. GPCRs have been shown to be hijacked by viruses as co-receptors for entry into host cells[43-45]. A French study reported that lower incidences of the symptomatic forms of COVID-19 were found among psychiatric patients (~4%) than clinical staff (~14%)[46]. An anti-psychiatric drug, chlorpromazine (**Figure 3**), has been repurposed for COVID-19 treatment, and is currently in phase III clinical trial (https://clinicaltrials.gov/ct2/show/NCT04366739). In another phase II clinical trial, fluvoxamine, a selective serotonin reuptake inhibitor (SSRI), was found to prevent more serious complications of COVID-19 infection (https://clinicaltrials.gov/ct2/show/NCT04342663). Pre-clinical data and clinical observations of those psychoactive drugs are promising; however, further clinical evidence and data are required to confirm the anti-SARS-CoV-2 effect of those psychoactive drugs.

In summary, we discovered that the autophagy and AP-1 signaling pathways might be potential targets for COVID-19 therapeutics through systematic mining of a large qHTS database. In addition, the class of neurology/psychiatry drugs was found significantly enriched with anti-SARS-CoV-2 active compounds, indicating that this class of drugs also has the potential to be repurposed as treatments for COVID-19 that warrant further investigation.

## Materials and Methods

*SARS-CoV-2 cytopathic effect (CPE) assay* Vero-E6 cells previously selected for high ACE2 expression [47] (grown in EMEM, 10% FBS, and 1% Penicillin/Streptomycin) were cultured in T175 flasks and passaged at 95% confluency. Cells were washed once with PBS and dissociated from the flask using TrypLE. Cells were counted prior to seeding. A CPE assay previously used to measure antiviral effects against SARS-CoV [48] was adapted for performance in 384 well plates to measure CPE of SARS CoV-2 with the following modifications. Cells, harvested and suspended at 160,000 cells/ml in MEM/1% PSG/1% HEPES supplemented 2% HI FBS, were batch inoculated with SARS CoV-2 (USA_WA1/2020) at M.O.I. of approximately 0.002 which resulted in approximately 5% cell viability 72 h post infection. Compound solutions in DMSO were acoustically dispensed into assay ready plates (ARPs) at 3 point 1:5 titrations. ARPs were stored at -20°C and shipped to BSL3 facility (Southern Research Institute, Birmingham, AL) for CPE assay. ARPs were brought to room temperature and 5µl of assay media was dispensed to all wells. The plates were transported into the BSL-3 facility were a 25 µL aliquot of virus inoculated cells (4000 Vero E6 cells/well) was added to each well in columns 3-24. The wells in columns 23-24 contained virus infected cells only (no compound treatment). A 25 µL aliquot of uninfected cells was added to columns 1-2 of each plate for the cell only (no virus) controls. After incubating plates at 37°C with 5% $CO_2$ and 90% humidity for 72 h, 30 µL of Cell Titer-Glo (Promega, Madison, WI) was added to each well. Following incubation at room temperature for 10 minutes the plates were sealed with a clear cover, surface decontaminated, and luminescence was read using a Perkin Elmer Envision (Waltham, MA) plate reader to measure cell viability.

*AP-1-bla ME-180 Assay* CellSensor® AP-1-bla ME-180 cell line and the culture medium components were purchased from ThermoFisher Scientific (Waltham, MA). These cells contain a beta-lactamase reporter gene under the control of AP-1 response element that has been stably integrated into ME-180 cells. Cells were cultured in DMEM medium supplemented with 10% dialyzed fetal bovine serum (FBS), 0.1 mM non-essential amino acids (NEAA), 1 mM sodium pyruvate, 25 mM HEPES, 100 U/ml penicillin, 100 µg/ml streptomycin, and 5 µg/ml of blastcidin at 37°C under a humidified atmosphere and 5% CO2. AP-1-bla ME-180 cells were used to screen the NPC compound collection. The positive controls, human epidermal growth factor (EGF) for AP-1-bla and tetraoctyl ammonium bromide for cytotoxicity assays, were purchased from Sigma-Aldrich (St. Louis, MO).

CellSensor® AP-1-bla ME-180 cells were suspended in 6 µL of assay medium (Opti-MEM with 0.5% dialyzed FBS, 0.1 mM NEAA, 1 mM sodium pyruvate, 100 U/ml penicillin, and 100 µg/ml streptomycin), and were dispensed at 2,500 cells per well in 1,536-well tissue culture treated black/clear bottom plates (Greiner Bio-One North America, NC) using a Multidrop Combi (Thermo Fisher Scientific). After incubation at 37°C for an overnight to facilitate cell adhesion, 23 nL of compounds and positive controls were transferred into the assay plates by a Pintool station (Wako Automation, San Diego, CA). The assay plates were incubated for 5 hr at 37°C. One µl of LiveBLAzer™ FRET B/G (CCF4-AM) substrate mix (Thermo Fisher Scientific) was added using an FRD and incubated at room temperature for 2 hr. The fluorescence signal was measured using an Envision plate reader (Perkin Elmer, Waltham, MA) at excitation 405 nm, and dual emissions at 460 and 530 nm. Data were expressed as relative fluorescence units (ratio of 460nm/530nm emissions). The cytotoxicity of each compound was tested in parallel in the same well by adding 4 µl/well of CellTiter-Glo reagent (Promega, Madison, WI) after beta-lactamase read into the assay plates using an FRD. After 30 min incubation at room temperature, the luminescence signal was measured using a ViewLux plate reader (Perkin Elmer). Cytotoxicity data were expressed as relative luminescence units.

**Table 1. Activity concordance between the SARS-CoV-2 CPE assay and top correlated assays**

| | Autophagy | AP-1 | Autophagy+AP-1 |
|---|---|---|---|
| TP | 88 | 84 | 118 |
| FN | 44 | 29 | 21 |
| FP | 142 | 399 | 473 |
| TN | 1887 | 1243 | 1682 |
| Sensitivity | 0.67 | 0.74 | 0.85 |
| Specificity | 0.93 | 0.76 | 0.78 |
| BA | 0.80 | 0.75 | 0.81 |

TP = True positive; number of compounds active in both the SARS-CoV-2 and the other assay

FN = False negative; number of compounds active in the SARS-CoV-2 assay but not active in the other assay

FP = False positive; number of compounds not active in the SARS-CoV-2 assay but active in the other assay

TN = True negative; number of compounds inactive in both assays

Sensitivity = TP/(TP+FN)

Specificity = TN/(TN+FP)

BA = Balanced accuracy; (sensitivity + specificity)/2

**Table 2. Potent (<20 µM) AP-1 compounds that were active in the SARS-CoV-2 CPE assay or reported as active in the literature.**

| Compound Name | AP-1 Potency (µM) | Autophagy Potency (µM) | SARS-CoV-2 CPE Potency (µM) | Literature Reported anti-SARS-CoV-2 | Neurology/ psychiatry |
|---|---|---|---|---|---|
| Oxyphenisatin | 0.22 | N/A | N/A | Y | |
| Clioquinol | 0.37 | >100 | 10.00 | | |
| Trimipramine maleate | 1.34 | 9.02 | 10.00 | | Y |
| Promethazine | 2.05 | 26.60 | 12.59 | Y | Y |
| Dimethisoquin | 2.49 | 10.12 | 12.59 | | Y |
| Cepharanthine | 3.79 | 13.33 | 2.00 | | |
| Pizotifen | 4.25 | 16.79 | 12.59 | | Y |
| Ethopropazine | 4.25 | >100 | 12.59 | | Y |
| Amitriptyline | 5.29 | 10.59 | 12.59 | | Y |
| Mefloquine | 5.78 | 29.85 | >100 | Y | |
| Tolterodine Tartrate | 6.49 | 14.96 | 12.59 | Y | |
| Tetraethylthiuram disulfide | 6.66 | 2.54 | >100 | Y | Y |
| Lynestrenol | 7.14 | >100 | 12.59 | | |
| Cyproheptadine | 7.28 | >100 | 3.98 | | |
| Benzydamine | 7.28 | >100 | 12.59 | | |
| Promazine | 7.56 | 5.69 | 10.00 | Y | Y |
| Triparanol | 7.56 | 21.13 | N/A | Y | |
| Fluphenazine | 7.56 | 13.33 | >100 | Y | Y |
| Imipramine | 7.76 | 13.33 | 8.91 | | Y |
| Loxapine succinate | 7.86 | 14.30 | 10.00 | | Y |
| Tripelennamine citrate | 8.49 | >100 | >100 | Y | |
| Homochlorcyclizine | 8.49 | 21.13 | 5.01 | | |
| Fluoxetine | 8.94 | 23.71 | 4.47 | | Y |
| Cyclobenzaprine | 9.52 | 14.96 | 12.59 | | Y |
| Chlorprothixene | 9.89 | 23.71 | 8.91 | | Y |
| Bepridil | 9.89 | >100 | 12.59 | | |
| Chlorpromazine | 10.15 | 16.79 | 11.22 | Y | Y |
| Difeterol | 11.10 | 29.85 | 12.59 | | |
| Zotepine | 11.10 | 13.33 | 12.59 | | Y |
| Spiperone | 11.99 | 13.33 | 12.59 | Y | Y |
| Clemastine fumarate | 11.99 | 23.71 | 11.22 | | |
| Maprotiline | 12.30 | 13.33 | 12.59 | | Y |
| Nylidrin | 13.45 | 26.60 | >100 | Y | |
| Duloxetine | 13.97 | 16.79 | 8.91 | | Y |
| Amoxapine | 14.34 | 5.96 | 8.91 | | Y |
| Clomipramine | 14.90 | 11.88 | 10.00 | Y | Y |
| Triflupromazine | 15.68 | 23.71 | 12.59 | | Y |
| Bromodiphenhydramine | 16.93 | N/A | 12.59 | | |
| Chlormadinone acetate | 17.59 | >100 | 12.59 | | |
| Tamoxifen citrate | 18.21 | 26.60 | 1.78 | Y | |
| Cyclomethycaine | 19.00 | N/A | 12.59 | | |
| Bencyclane | 19.74 | N/A | 12.59 | | |

(a)

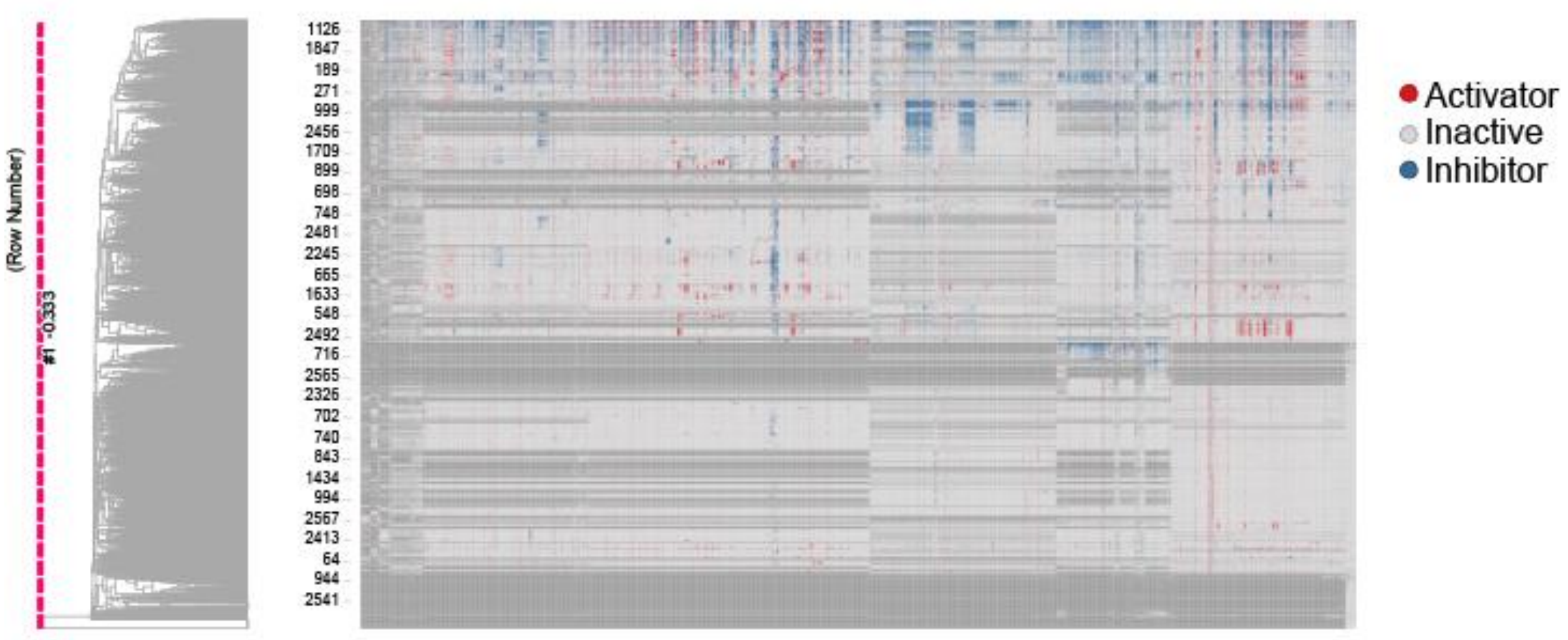

(b)

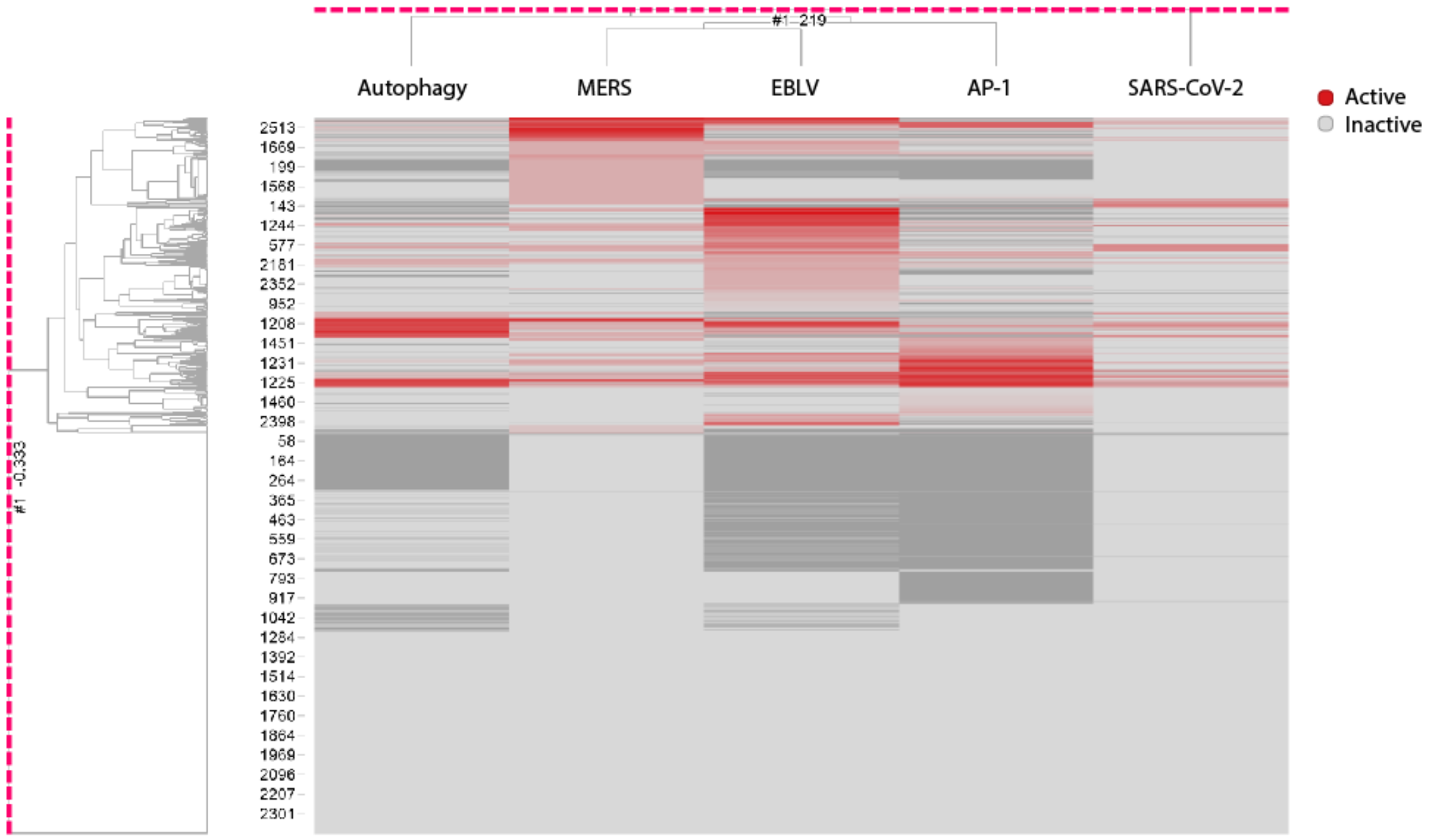

**Figure 1**. Compound activity profiles from NPC screens. In the heat map, each row is a compound and each column is an assay readout. The heat map is colored by "curve rank",[22,23] a numeric measure

(between -9 and 9) of compound activity based on potency, efficacy, and the quality of the concentration response curve, such that a large positive number indicates a strong activator (red), a large negative number indicates a strong inhibitor (blue), and 0 means inactive (light gray). Dark gray indicates missing data. (a) All assays. (b) Assays most correlated with SARS-CoV-2.

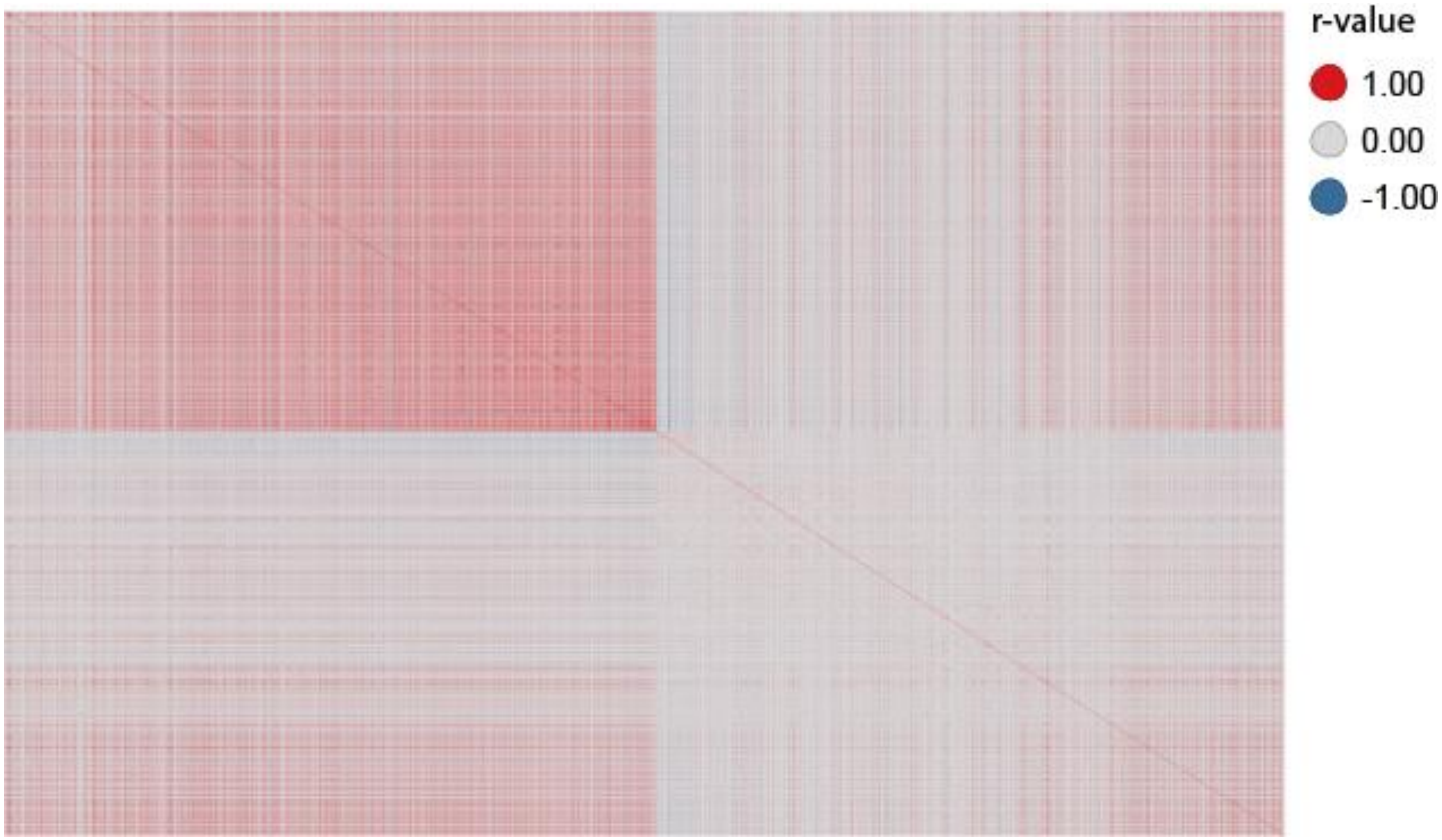

**Figure 2**. Activity profile correlations between assays. In the heat map, each row/column is a different assay readout. The heat map is colored by the correlation coefficient (*r*), such that darker shades of red indicate stronger positive correlations and darker shades of blue indicate stronger negative correlations. Gray means no correlation found.

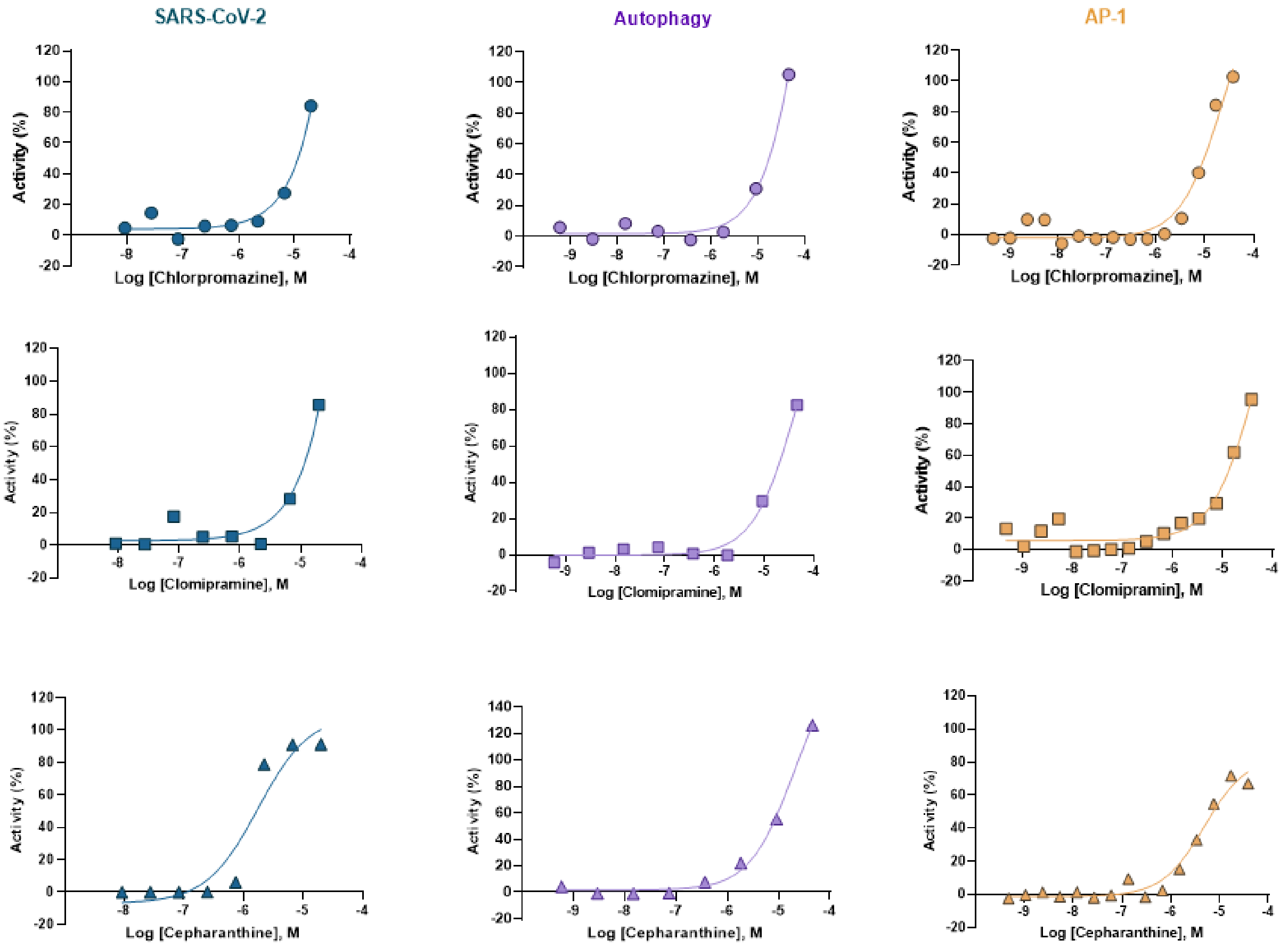

**Figure 3**. Example concentration-response curves of compounds active in the SARS-CoV-2, autophagy, and AP-1 assays. Chlorpromazine is a tricyclic antipsychotic, clomipramine is a tricyclic antidepressant, and cepharanthine is a natural product anti-inflammatory that is approved in Japan.[49]

| Drug class | Total compounds | Activity in anti-COVID19 assay | | Enrichment (fold) | *P* value |
|---|---|---|---|---|---|
| | | Class active rate | Background active rate | | |
| Neurology/Psychiatry | 359 | 0.21 | 0.08 | 2.56 | 2.41E-11 |
| Infectious disease | 315 | 0.11 | 0.10 | 1.09 | 0.61 |
| Cardiology | 209 | 0.14 | 0.10 | 1.44 | 0.05 |
| Endocrinology | 129 | 0.07 | 0.10 | 0.69 | 0.29 |
| Gastroenterology | 140 | 0.14 | 0.10 | 1.47 | 0.08 |
| Oncology | 95 | 0.09 | 0.10 | 0.95 | 1.00 |

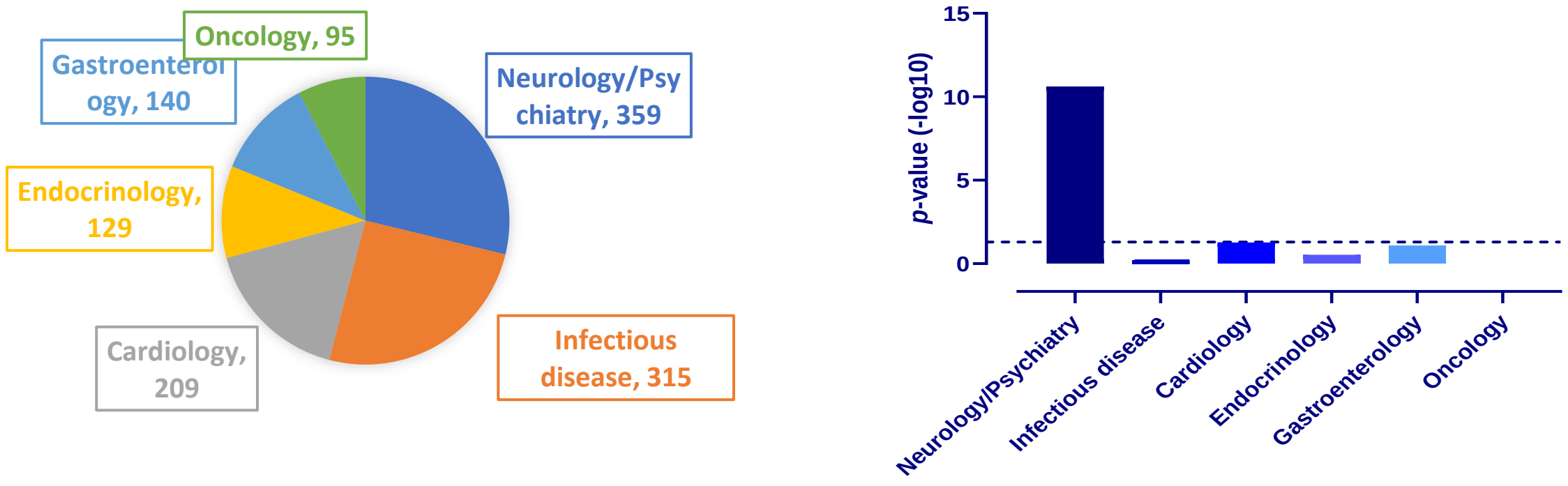

Figure 4. Enrichment of neurology/psychiatry drugs in anti-SARS-CoV-2 active compounds. Top panel, activity statistics of six common drug classes in the SARS-CoV-2 CPE assay. Bottom left, distribution of the six common drug classes in the NPC library. Bottom right, significance of enrichment of anti-SARS-CoV-2 actives in each drug class. Only the class of neurology/psychiatric drugs was significantly enriched. The dotted line indicates the threshold for statistical significance.

## Acknowledgements

This work was supported by the Intramural Research Programs of the National Center for Advancing Translational Sciences, National Institutes of Health. The authors would like to thank Hui Guo, Xin Hu, and Min Shen for assistance with CPE assay data processing, and Richard Eastman, Zina Itkin, and Paul Shinn for compound management and plating.

## Author Contributions

R.H. and H.Z. conceived the research and designed the study. C.Z.C. and S.S. performed the experiments, collected data and aided data interpretation. R.H. performed statistical analysis of all data. H.Z. aided data analysis and visualization. R.H., H.Z. and M.D.H. wrote the manuscript. R.H., M.D.H., M.X., W.Z. and A.S. directed the research. All authors reviewed the manuscript.

## Competing financial interests

The authors declare no competing financial interests.